\documentclass[a4paper,11pt]{article}
\pdfoutput=1 

\usepackage{jheppub} 
\graphicspath{{Figures/}}
\newcommand{\beq}{\begin{equation}}
\newcommand{\eeq}{\end{equation}}
\let\p=\partial
\let\Om=\Omega
\let\t=\tilde

\title{Page curves for a family of exactly solvable evaporating black holes}
\author[a]{Xuanhua Wang,}
\author[b,c,*]{Ran Li,}
\author[a,b,*]{Jin Wang \note[*]{Corresponding author}}
\affiliation[a]{Department of Physics and Astronomy, Stony Brook University, Stony Brook, NY 11794, USA,}
\affiliation[b]{Department of Chemistry, Stony Brook University, Stony Brook, NY 11794, USA}
\affiliation[c]{Department of Physics, Henan Normal University,
 Xinxiang 453007, China}
\emailAdd{xuanhua.wang@stonybrook.edu}
\emailAdd{liran@htu.edu.cn}
\emailAdd{jin.wang.1@stonybrook.edu}

\abstract{We study the entanglement entropy of a one-parameter family of exactly solvable gravities in the 2-dimensional asymptotically-flat space. The islands and Page curves of eternal, evaporating and bath-removed black holes are investigated. The different theories in this parameter class are identified through field redefinitions which leave the island invariant. The Page transition is found to occur at the first a third of the black hole life time in the evaporating case for this family of solutions. In addition, we consider gluing the equilibrium black hole and the evaporating one along a null trajectory and study the effect of gluing on the islands and Page curves. In the glued space, the island jumps across two different geometries at a certain retarded time. As a result, the Page transition is stretched and split into two separate ones---the first transition happens when the net entropy generation stops and the second one occurs as the early radiation effectively starts to become purified. Finally, we discuss the issues concerning the inconsistent rates of purification and the paradox related to the state of the radiation.}

\begin{document} 
\maketitle


\section{Introduction and motivation}
The issues concerning the purity of Hawking radiation has been around for decades \cite{Hawking:1974sw,Page:1993wv}. Recent breakthrough was made through the gravitational path integral of the replica wormholes geometries resulting in the formulation of the quantum extremal surface \cite{Almheiri:2019hni,Almheiri:2019yqk,Almheiri:2019qdq,Almheiri:2019psf,Penington:2019npb,Engelhardt:2014gca,Penington:2019kki,Engelhardt:2013tra}. This allows one to calculate the full quantum description of the fine-grained entropy of a gravitational system in the large-$N$ limit. Several models in the asymptotically AdS space and asymptotically flat space have been explicitly studied using the approach of quantum extremal surfaces, and are shown to reproduce the corresponding Page curves \cite{Almheiri:2019psf,Hollowood:2020cou,Balasubramanian:2020xqf,Hartman:2020swn,Gautason:2020tmk,Miyata:2021ncm}. This is an indication of the unitarity in the process of Hawking radiation and the consistency with one of the basic principles of quantum mechanics. 

The conservation of information is one of the basic principles of quantum mechanics. However, it is at odds with Hawking's calculation of the black hole radiation \cite{Hawking:1974sw,Hawking:2005kf}. The black hole information problem in the $(3+1)$-dimensional spacetime gets complications from the grey body factor in 4 dimensions in addition to the fact that most 4-dimensional systems are not exactly solvable \cite{Penington:2019npb,Krishnan:2020oun}. The existing studies on the black hole information in four dimensions mostly concentrate on the equivalent description of the information paradox of the eternal black holes in thermal baths and the black holes in the braneworld in doubly holographic models \cite{Almheiri:2019psy,Wang:2021woy,Hashimoto:2020cas,Chen:2019uhq,Rozali:2019day,Karananas:2020fwx,Alishahiha:2020qza,Geng:2020fxl,Bhattacharya:2020uun,Bhattacharya:2020ymw,Krishnan:2020fer}. The best-understood case regarding the entanglement entropy in the dynamical black hole evaporation process is the JT gravity. The JT gravity has the advantage of being embedded in the asymptotically AdS space and the technology of the AdS/CFT duality is easily applicable \cite{Teitelboim:1983ux,Jackiw:1984je,NavarroSalas:1999up,Chen:2020jvn,Anderson:2021vof,Chen:2020uac,Chen:2020hmv,Hernandez:2020nem,Li:2020ceg}. However, very few solvable evaporating models have been explicitly calculated. Exactly solvable models besides the JT gravity, in particular those embedded in asymptotically flat spaces and more general spaces which better resemble the black holes in our universe are worth of careful studies and may eventually shed light on the nature of the islands and the quantum information of gravity \cite{Gautason:2020tmk,Hartman:2020swn,Miyata:2021ncm}.

The 2-dimensional models are of particular interest due to several reasons. First of all, the (1+1)-dimensional models introduced by Callan, Giddings, Harvey and Strominger (CGHS) and some of its variants present some essential features of the black holes in 4D such as the singularity and the event (apparent) horizon \cite{Callan:1992rs,Grumiller:2002nm}. Furthermore, in the large-N limit, these models incorporate the Hawking radiation as well as its backreaction on the black hole geometry and yet still remain exactly solvable. This allows us to study the full dynamical process of the black hole formation and in particular the evaporation analytically. For a nice review of 2D black holes, see Refs. \cite{Strominger:1994tn,Grumiller:2002nm}. 

The details of how the Hawking radiation reacts back on the black hole geometry are necessary for understanding the information of an evaporating black hole but hard to obtain analytically. The CGHS model offers a simplified recourse to the problems in (1+1)D and admits classical black hole solutions. The original proposal suggest that the singularity and the event horizon of the black hole are removed by the quantum correction of the matter fields. This issue of failing to form a black hole was rectified by Russo, Susskind and Thorlacius (RST) \cite{Russo:1992ht,Russo:1992ax}. The RST model allows the formation of a black hole with a singularity and an event horizon in the large-N limit. In this model, solutions cannot be continued past the singular region and admit boundaries of the spacetime.

Another 2D dilaton model which is exactly solvable was found by Bose, Parker and Peleg (BPP) \cite{Bose:1995pz}. The BPP model admits black hole solutions and is another modification of the CGHS model. The endpoint of an evaporating BPP black hole can be continuously connected to a unique vacuum solution after a thunderbolt of energy emanating from the intersecting point. The resulting vacuum geometry has an infinite throat and is geodesic complete which differs from other 2D models. Nevertheless, it is noticed that both the RST model and the BPP model belong to the same one-parameter family of 2d gravities which was shown to admit exact solutions \cite{Cruz:1997nj,Cruz:1995zt,Cruz:1996pg}. 

In this paper, we apply the quantum extremal surface to the study of the dynamical RST-BPP black holes and calculate the entanglement entropy of the black holes in various circumstances. The gravitational entropy formula was proposed using the holographic argument, and was later derived from the gravitational path integral without holography \cite{Almheiri:2019qdq,Lewkowycz:2013nqa,Penington:2019kki,Dong:2016hjy}. The entanglement entropy with the island is given by
\begin{equation}
\label{GEformula}
 S(R)  =  \min \left\{\mathrm{ext}\left[ \frac{\mathrm{Area}(\partial I)}{4G_{\rm N}}  + S_{\rm matter}(R\cup I) \right]\right\} \,,
\end{equation}
where $I$ is the island, $R$ is the radiation, and the expression inside the extremization and the minimization operators is called the {\it generalized entropy} $S_{\rm gen}$ \cite{Almheiri:2019hni,Engelhardt:2014gca,Faulkner:2013ana,Engelhardt:2013tra,Ryu:2006ef,Ryu:2006bv}. The entanglement entropy incorporating the island has been applied to the systems far beyond the asymptotically AdS space to the asymptotically flat space, de-Sitter space and cosmologies, etc \cite{Hartman:2020khs,Balasubramanian:2020coy,Almheiri:2020cfm,Geng:2021wcq,Marolf:2020rpm,Raju:2020smc}. We study the configuration of the islands in the eternal black holes, the dynamical black holes formed by collapsing matter shells and the bath-removed evaporating black holes. We show that corresponding Page curves are recovered for each of the three cases. 

Similar discussion of the islands in the RST model can be found in previous papers for the case of eternal and evaporating black hole \cite{Gautason:2020tmk,Hartman:2020swn}, while the BPP model has not been discussed. Both the RST and the BPP model are points in the trajectory of the one-parameter family and it is unclear if the conclusion drawn from RST can be extended into a larger class of exactly-solvable models. 
In this paper, we show that the result for the entanglement entropy does not depend on the free parameter in the leading order at late times and large distances, and thus the evaporating RST Page curves is extended into a larger class. How to understand the Page transition has been an important question especially after the Page curve was calculated.  Another focus of this paper is the gluing of two geometries, where we cut the Penrose diagram of an equilibrium black hole along a light-like trajectory and glue it with the evaporating geometry. The gluing of the two geometries has an effect similar to an injection of negative energy on top of the stable equilibrium black hole background. In this case, we find that the island will jump from outside the horizon to the inside of the horizon and the Page curve may have a plateau depending on the time of occurrence of the gluing. The ``plateaued" Page curve indicates two different Page transitions. The Page transition in an ordinary evaporating or eternal black hole scenario is separated into two separate transitions in this model. It is unclear how to interpret the two Page times beyond the entanglement islands or the physical implications of the discontinuity inside the black hole. One possible explanation is using the ER=EPR argument but this argument raises concerns about the ambiguity of the state of the radiation \cite{Akers:2019nfi,Bousso:2020kmy}.

The paper is organized as follows. In section two, we introduce the RST-BPP model and the CFT coupling. We discuss the general solutions of the model and the quantum corrected area in the black hole solutions. In section three, we identify the islands in the eternal black hole and the evaporating black hole formed by collapsing matter shells. We show that the Page transition for the evaporating black hole occurs at a third of the evaporation time. In section four, we investigate the evaporating black holes which are initially in equilibrium with incoming radiations emitted from external sources. We quantify the island configurations and the Page curves when the bath is removed at different stages in the evaporation process. The discussions of our results and issues regarding the rates of purification in the glued and evaporating geometries are presented in the last section.

\section{RST-BPP model}
The theories admitting special conformal symmetries are shown to be those having a classical exponential potential in the curved spacetime \cite{Cruz:1997nj}
\beq S_{0}=\frac{1}{2\pi}\int d^2x \sqrt{-g}(R \phi+4\lambda^2e^{2\beta\phi})\,, \label{actn:conf}
\eeq
where $\phi$ is the dilaton field and $\lambda$ is a free parameter characterizing the relic of the scalar curvature tangent to the two-sphere. One particular example in this category of theories is the classical CGHS model. The CGHS action which is given by
\beq
S_{\rm CGHS}=\frac{1}{2\pi}\int d^2x \sqrt{-g}e^{-2\phi}(R + 4(\nabla\phi)^2 + 4\lambda^2)
\eeq
can be recovered by setting $\beta=0$ \footnote{Here $\beta$ is a free parameter describing the strength of the exponential coupling of the dilaton field not the inverse temperature.} in Eq.~\eqref{actn:conf} and redefining the fields $g_{\mu\nu}=\overline{g}_{\mu\nu}e^{-2\bar{\phi}},\ \phi=e^{-2\bar{\phi}}$. This model can be regarded as the low-energy effective action that governs the radial modes propagating on the near-extreme magnetically charged black hole of four-dimensional dilaton gravity \cite{Fiola:1994ir}. In this paper, we discuss a family of exactly-solvable models which interpolating between the BPP and RST models
\beq
S=S_{\rm CGHS}+\frac{N}{24\pi}\int d^2x \sqrt{-g}\left[(1-2a)(\nabla\phi)^2+(a-1)\phi R)\right]\,.\label{S}
\eeq
The RST model is recovered by setting $a=\frac{1}{2}$ and the BPP model can be obtained by requiring $a=0$. One distinct property of this family of models is that they can be exactly solved under the semi-classical approximation \cite{Cruz:1995zt}. 

\subsection{CFT coupling and entanglement entropy}
The above Lagrangian and the solutions represent a classical black hole in the vacuum with no additional coupling to the other quantum fields. Here we consider the case when the classical gravity is coupled to a quantum field theory with conformal symmetry. The Lagrangian of the CFT is given by
\beq
S_{\rm CFT}=-\sum_{k=1}^N\frac{1}{4\pi}\int d^2x \sqrt{-g}(\nabla f_k)^2\,,
\eeq
where $f_k$ are a set of $N$ massless matter fields. 

The inclusion of the matter CFT results in the conformal anomaly in a curved spacetime. The trace of the energy-momentum tensor which is zero classically becomes nonzero if the one-loop correction is considered. The expectation value of the trace of the energy-momentum tensor is proportional to the Ricci scalar,
\beq
\langle T_\mu^\mu \rangle \equiv \langle T \rangle =\frac{N}{24} R\,,
\eeq
where $R$ here denotes the Ricci scalar. The backreaction of the CFT and the conformal anomaly can be incorporated into the equations of motion by including in the Lagrangian an additional non-local term,
\beq
S_P=-\frac{N}{96\pi}\int d^2 x\sqrt{-g}R\ \Box^{-1}R\,,\label{S_P}
\eeq
which is called the Polyakov action.

The light-cone coordinates are defined as
\begin{align}
    x^\pm=x^0\pm x^1\,.
\end{align}
For the two-dimensional gravity in the conformal gauge, i.e. $d^2s=-e^{2\rho}dx^+dx^-$, the Ricci scalar is given by \beq R=8e^{-2\rho}\partial_+\partial_-\rho\,. \eeq
The connection coefficients are $\Gamma_{++}^+=2\partial_+\rho$ and $\Gamma_{--}^-=2\partial_-\rho$. The trace of the energy-momentum tensor is $T=-4e^{-2\rho}T_{+-}$, therefore, for N scalars we have
\beq
\langle T_{+-}^f \rangle =-\frac{N}{12}\partial_+\partial_-\rho\,.
\eeq
The covariant derivatives of the energy-momentum tensor are zero due to the energy-momentum conservation in general spacetimes, and this gives
\beq
\begin{split}
    \p_+T_{--}+\p_-T_{+-}-\Gamma_{--}^-T_{+-}=0\,,\\
    \p_+T_{-+}+\p_-T_{++}-\Gamma_{++}^+T_{-+}=0\,.
\end{split}
\eeq
Integrating the equations returns the one-loop contribution to the diagonal components of the stress tensor,
\begin{align}
    &\langle T_{++}^f \rangle =-\frac{N}{12}(\partial_+\rho\partial_+\rho-\partial_+^2\rho+t_+(\sigma^+))\,,\nonumber\\
    &\langle T_{--}^f \rangle =-\frac{N}{12}(\partial_-\rho\partial_-\rho-\partial_-^2\rho+t_-(\sigma^-))\,.
\end{align}
Here, $\langle T_{++}^f \rangle$ and $\langle T_{--}^f \rangle$ are the vacuum expectation value of the stress tensor of the $f$-waves. $t_\pm(\sigma^\pm)$ are the vacuum energy due to different boundary conditions and are coordinate-dependent. $t_\pm(\sigma^\pm)$ are determined by requiring that the stress tensor vanishes in the dilaton vacuum and along the past null infinity. The first two terms are derivatives of the conformal factor of the spacetime which can be ignored in the asymptotically flat region and the stress tensor in the flat metric is
\begin{align}
    \left. \langle T_{\pm\pm}^f(\sigma^\pm)\rangle\right|_{\rm boundary}= -\frac{N}{12}t_\pm\,.
\end{align}
The conformal reparametrization $\sigma^\pm \rightarrow y^\pm$ transforms the conformal factor as 
\beq
\rho(y^\pm)=\rho(\sigma^\pm)-\frac{1}{2}\log \frac{dy^+}{d\sigma^+}\frac{dy^-}{d\sigma^-}\,.
\eeq
The tensor transformation of the energy-momentum tensor $\langle T_{\pm\pm}^f \rangle$ leads to the anomalous transformation of the boundary term $t_\pm$ under a Weyl rescaling of the metric,
\beq
\left(\frac{dy^\pm}{d\sigma^\pm}\right)^2t_{\pm}(y^\pm)=t_{\pm}(\sigma^\pm)+\frac{1}{2}\{y^\pm,\sigma^\pm\}\,.\label{t_transform}
\eeq
where $\{y,\sigma\}$ is the Schwarzian derivative defined as 
\begin{align}
    \{y,\sigma\}=\frac{y'''}{y'}-\frac{3}{2}\frac{(y'')^2}{(y')^2}\,.
\end{align} 

To calculate the generalized entropy of the conformal field using the CFT results, we need to transform the CFT into the vacuum state with vanishing stress tensors. 

It can be shown that the entanglement entropy $S_A=\frac{c}{3}\log\frac{l}{\epsilon}$ for CFT$_2$. For d-dimensional theories it is proportional to the area of the $d-2$-dimensional hypersurface up to lower-order divergent (or constant) terms \cite{Hertzberg:2010uv}. A simpler way to calculate the entanglement entropy is through twist operators on an orbifold assuming the replica symmetry \cite{Calabrese:2004eu}. The result is the following: the ratio of partition functions is the same as the correlation function arising from the insertion of primary scaling operators $\phi_n$ with scaling dimension $\Delta_n = (c/12)(1-1/n^2)$, into each of the $n$ (disconnected) sheets. 
\begin{align}
    {\rm tr_A} \rho_A^n=\Pi_{k=0}^{n-1}\langle\mathcal{T}_k(x_1,0)\mathcal{T}_k(x_2,0)\rangle =|x_1-x_2|^{-2n \Delta_n}\,.
  \end{align}
In curved spacetime, the conformal invariance of CFT gives 
\begin{align}
    \langle\mathcal{T}(x_1,0)\mathcal{T}(x_2,0)\rangle_{\Om^{-2}g}=\Om^{\Delta_n}(x_1,0)\Om^{\Delta_n}(x_2,0)\langle\mathcal{T}(x_1,0)\mathcal{T}(x_2,0)\rangle_g\,
\end{align}
under the conformal transformation $g\rightarrow\Om^{-2}g$. Therefore, the von Neumann entropy in the Wyle transformed space and the flat space are related by
\begin{align}
    S_g = S_{\rm flat}+\frac{c}{6}\sum_{x_i}\log \Omega(x_i)\,.
\end{align}

\subsection{Solutions of geometry}
The RST-BPP models are exactly solved in Refs.~\cite{Cruz:1995zt,Cruz:1996pg}. The equations of motion can be obtained by varying the total action given by the sum of Eq.~\eqref{S} and Eq.~\eqref{S_P} and imposing gauge conditions. After choosing the conformal coordinates $d^2s=-e^{2\rho}dx^+dx^-$ and gauge condition $\rho=\phi$, the equations of motion are 
\begin{gather}
    \p_+\p_-(e^{-2\phi}+\frac{Na}{12}\phi)+\lambda^2=0\,,\\
    \p_\pm^2\left(e^{-2\phi}+\frac{Na}{12}\phi\right)+T_{\pm\pm}^{cl}+\frac{N}{12}t_\pm=0\,,
\end{gather}
where $T^{cl}_{\pm\pm}$ is the classical stress tensor which transforms covariantly in both indices, and $t_\pm(x^\pm)$ are the boundary terms from the Polyakov action which transform anomalously. The whole class of solutions can be identified after the field redefinition and the interpolation parameter can be absorbed into the redefined variable. To be specific, we set $\lambda=1$ and define the ``Kruskal variables" as 
\begin{gather}
    \phi_k=\phi+\frac{1}{2}\log\frac{N}{12}\,,\\
    \Omega_k=e^{-2\phi_k}+a \phi_k-\frac{a}{2}(1-\log \frac{a}{2})+\frac{1}{4}\,, \label{Def:Omega}
\end{gather}
where $\Om_k$ corresponds to the area of a static spherical surface in the 2D theory. The constant is chosen such that the causal structure of the spacetime is independent of the parameter $a$ and it admits the same boundary as in the dilaton vacuum solution. We stick to this convention through out this paper and will drop the subscript ``$k$" on $\phi$ and $\Omega$. For the BPP black hole, $\Om$ should be understood as the limit of $a\rightarrow 0$. With the above redefinition, the equations of motion then read
\begin{gather}
    \p_\pm^2\Omega+T_{\pm\pm}^{cl}+t_\pm=0\,,\quad    \p_+\p_-\Omega+1=0\,, \label{Eq:eom}
\end{gather}
where we have rescaled the stress tensor $T_{\pm\pm}^{cl}$ by a factor of $N/12$ for notational convenience. 

Curvature singularities appear when $R=8e^{-2\rho}\partial_+\partial_-\rho$ diverges. In the Kruskal coordinates, one finds
\begin{align}
    \p_+\p_-\rho=\frac{1}{\Om'}\left[\p_+\p_-\Om-\frac{\Om''}{\Om'^2}\p_+\Om\p_-\Om\right]\,,
\end{align}
where $\Om'\equiv \frac{d\Om}{d\rho}=\frac{d\Om}{d\phi}$. The singular region is located at the curve $\frac{d\Om}{d\phi}=0$, which gives the boundary of the spacetime $\Om=\frac{1}{4}$ according to Eq.~\eqref{Def:Omega}. This boundary condition must be satisfied for all the solutions discussed in this study. 

From Eq.~\eqref{Eq:eom}, we can find the general solution to the equations of motion in the absence of matter ($T_{\pm\pm}^{cl}=0$)
\begin{align}
    \Omega=-x^+x^- -\int^{x^+}\int^{x^+}t_+(x^+) -\int^{x^-}\int^{x^-}t_-(x^-)+bx^++cx^-+M\,,\label{Eq:Omega}
\end{align}
where $M$ is related to the ADM mass by $M=\frac{12\pi}{N}M_{\rm ADM}$. 

For example, the asymptotically Minkowski vacuum solution can be obtained by requiring $b=c=M=0$ and that the energy flux in the asymptotically flat coordinates $\sigma^\pm$ defined by $x^\pm=\pm e^{\pm \sigma^\pm}$ vanishes identically, i.e. $t_\pm(\sigma^\pm)=0$. Transforming to the $x^\pm$ coordinates according to the transformation rule Eq.~\eqref{t_transform} gives
\begin{align}
    t_\pm(x^\pm)=-\frac{1}{4(x^\pm)^2}\,.
\end{align}
Setting the vacuum energy condition in Eq.~\eqref{Eq:Omega}, we obtain the geometry of the vacuum solution 
\begin{align}
    \Om=-x^+x^--\frac{1}{4}\log(-x^+x^-)\,.
\end{align}
This solution does not admit radiation as indicated by $t_\pm(\sigma^\pm)=0$. Note that spacetime we are interested in is bounded by $x^+x^-<-\frac{1}{4}$ and $\pm x^\pm>0$ which is required by Eq.~\eqref{Def:Omega}. The boundary of the spacetime is where the dilaton coupling $e^{2\phi}$ becomes strong and the semi-classical approximation breaks down. Near those regions, the quantum gravitational correction becomes large and we do not pretend to have the full quantum mechanical description of it. In the coordinates $\sigma^\pm$ , the geometry is 
\begin{align}
    \Om=e^{-2\phi}+a \phi+\rm{constant}=e^{2\sigma}-\frac{1}{2}\sigma\,,
\end{align}
where $ \sigma\equiv\frac{\sigma^+-\sigma^-}{2}$. In the region far away from the boundary points ($\sigma\rightarrow\infty$), $\phi\approx-\sigma$ and the metric is 
\begin{align}
    ds^2=-e^{2\rho-2\sigma}d\sigma^+d\sigma^-\approx -d\sigma^+d\sigma^-\,,
\end{align}
which is the linear dilaton vacuum. When $a=\frac{1}{2}$, the solution is exactly Minkowski space as $ds^2=-d\sigma^+d\sigma^-$, otherwise it is asymptotically flat in the region far away from the boundary.

\subsection{Quantum-corrected Bekenstein-Hawking entropy}
The coarse-grained entropy of the RST-BPP black holes incorporating the next-to-leading order corrections can be calculated following the thermodynamic argument given in Ref.~\cite{Fiola:1994ir}. For a cloud of photon gas near the apparent horizon, the falling of the matter into the horizon will shift the position of the horizon and consequently change the entropy of the black hole. Incorporating the total entropy change and the entropy change of the radiation, we obtain the entropy change of the black hole from the second law of thermodynamics and thus the ``corrected area" of the black hole.

For the RST-BPP black hole, the temperature $T_{\rm BH}=1/2\pi$ is constant. The infinite specific heat of the black hole results in a fluctuation in the thermal equilibrium which can be suppressed by approximately $1/\sqrt{N}$ in the large $N$ limit. We assume this limit and the validity of the equilibrium thermodynamics. The classical limit of the black hole entropy is 
\beq
S_{\rm BH}=2e^{-2\phi_H}\,,
\eeq
where $\phi_H$ denotes the value of the dilaton field at the apparent horizon. 

For a black hole in the Hartle-Hawking state, we consider adding a coherent state of left-moving matter matter falling into the black hole and the solution in Kruskal gauge takes the form, 
\beq
\Omega=-x^+(x^-+P_+(x^+))+M(x^+)\,,
\eeq
where $P_+$ is the total incoming Kruskal momentum. The solutions of the horizon is given by $x_H^-(x^+)=-P_+(x^+)$ and $\Omega_H=M(x^+)$. From the second law of thermodynamics, the total entropy change of this process is 
\beq \Delta S_{\rm total}=\Delta S_{\rm BH}+\Delta S_{\rm matter}=\int dM_{\rm ADM}/T=\frac{N}{12\pi T}\Delta M =\frac{N}{6}\Delta \Omega_H\,,\eeq
where $M_{\rm ADM}=\frac{N}{12\pi}M$ is the ADM mass. As is given by Ref.~\cite{Fiola:1994ir}, the matter part of the entropy change is $\Delta S_{\rm matter}=\frac{N}{6}\Delta\phi_H$. From the above equations, we can easily derive the entropy change of the black hole. Integrating this equation leads to the entropy of the black hole up to a constant\footnote{The constant can be determined by matching the zero entropy with the singularity point which does not concern us here.}. In Kruskal coordinate, the result is
\beq
S_{\rm BH}=\frac{N}{6}(\Omega_H-\rho_H)+\rm constant \,, \label{Eq:2.last}
\eeq
and this serves as the next-to-landing corrected area for calculating the Bekenstein-Hawking entropy.

\section{Eternal and evaporating RST-BPP black holes}
The general methods for calculating the generalized entropy in asymptotically-flat 2D gravities are similar. For this section on eternal and evaporating black holes, readers unfamaliar with this model can refer to the calculation of the islands in eternal black holes and evaporating black holes in the RST model \cite{Gautason:2020tmk,Hartman:2020swn}. The conclusion is summarized as follows---the island configuration is invariant in the whole parameter class. The difference only appears in the CFT part due to the different metrics. In general, the free parameter which interpolates between different models has a negative contribution to the generalized entropy. However, this contribution is exponentially suppressed in the late time and large distance limit, and therefore the Page curves is invariant in the one-parameter family to the leading order.

\subsection{Islands in eternal black holes}

The eternal black hole solution can be calculated by letting $b=c=t_\pm=0$, which gives
\beq
\Om=-x^+x^-+M\,.
\eeq
To ensure that the singularity lies inside the apparent horizon, we require that $M>\frac{1}{4}$. Because $\Omega$ is the area of the black hole, the location of the apparent horizon is obtained by requiring $\p_+\Om=0$. This equation is equivalent to that following an outgoing light ray, the area stays constant (is not expanding or is just about to shrink), which is the definition of the apparent horizon. For an eternal black hole, the location of the apparent horizon and the event horizon coincides at $x_+x_-=0$. It is easy to check that the energy flux due to the vacuum Casimir energy is
\beq
\p_\pm^2\Omega=-t_\pm=0\,.
\eeq
Near the future null infinity, $-x^+x^-\gg M$. In this case, we return to the asymptotically flat coordinates $x^\pm=\pm e^{\pm \sigma^\pm}$. The metric is flat near the asymptotic null infinity $ds^2\simeq -d\sigma^+ d\sigma^-$. Transforming the energy flux $t_\pm(x^\pm)$ to the flat coordinates gives
\beq
t_\pm(\sigma^\pm)=\frac{1}{4}\,.
\eeq
This suggests that the black hole has an equivalent amount of incoming and outgoing flux which is the feature of the eternal black hole in equilibrium with the bath. The energy flux gives the Hawking temperature 
\beq
T_{\rm BH}=\frac{1}{2\pi}\,,
\eeq
which is a constant during the evaporation.

\begin{figure}[t]
\centering
	\includegraphics[width=90mm]{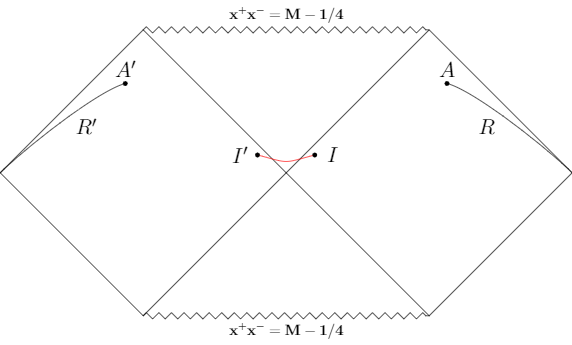}
	\caption{Penrose diagram for an eternal dilaton black hole. The island is between the boundary points $I'$ and $I$. The radiation is collected beyond the cutoff surface at $A'$ and $A$ to the spacial infinity as indicated by $R$ and $R'$.}
	\label{fig:eternal}
\end{figure}

For the eternal black hole, the geometry is given by 
\begin{align}
    \Om_a=-x^+x^-+M=e^{-2\rho}+a\rho+\rm constant\,.
\end{align}
Assuming that the anchor points are far from the horizons, the conformal factor at point $``a"$ can be calculated to be approximately as follows
\begin{align}
    \rho_a=-\frac{1}{2}\log(-x^+x^-+M+\frac{a}{2}\log(-x^+x^-+M))+({\rm higher\ orders})\,.
\end{align}
For eternal black holes, $t_\pm(x^\pm)=0$ and we use the coordinates to calculate the vacuum entanglement entropy. In the holographic calculation, the entropy without island can be treated as a special case of the island configuration where the island shrinks to zero. Without considering the configuration of the islands, the entropy is given by 
\begin{align}
    S&=\frac{N}{6}\log\left[d_{\rm flat}^2(A',A)e^{\rho(A')}e^{\rho(A)}\right]\nonumber\\
    &=\frac{N}{12}\log\left[\frac{(x_2^+-x_1^+)^2(x_2^--x_1^-)^2}{\left(-x_1^+x_1^-+M+\frac{a}{2}(-x_1^+x_1^-+M)\right)\left(-x_2^+x_2^-+M+\frac{a}{2}(-x_2^+x_2^-+M)\right)}\right]\,,
\end{align}
where $x_1^\pm$ are coordinates of endpoint $A'$ and $x_2^\pm$ are coordinates for anchor point. $\rho(A)$ is the value of the conformal factor $\rho$ at the point A. Note that only the vacuum CFT is included with no island contribution, and this gives the bulk entropy. The coordinates of the anchor points are related by $x_1^\pm=x_2^\mp$. Then we transform to the asymptotically flat coordinates $(t,y)$, where
\begin{align}
    x^+=e^{t+y}\,,\quad x^-=-e^{-t+y}\,.
\end{align}
Expressed in the coordinates of anchor point $A$, we have the generalized entropy approximately as follows:
\begin{align}
    S=\frac{N}{3}t+\frac{N}{3}y+\frac{N}{6}\log\left(\frac{1}{M+e^{2y}+\frac{a}{2}\log(M+e^{2y})}\right) \,,
\end{align}
where $t$ and $y$ are the time and spacial coordinates of $A$. At late times and large distances, the entropy grows approximately linearly with time $S \sim \frac{N}{3}t$. The free parameter $a$, which contributes negatively to the generalized entropy, is exponentially suppressed in the late times and large distances. Therefore, the qualitative feature of the Page curve is independent of the $``a"$ and diverge at late times The fine-grained entropy of the black hole eventually exceeds (twice of) the Bekenstein-Hawking entropy. 

Assuming the island configuration as shown in Fig.~\ref{fig:eternal}. In the late times, the cross terms in the bulk entropy decays exponentially. The generalized entropy, which is the Bekenstein-Hawking entropy of the island plus the bulk fine-grained entropy of the CFT, is approximately 
\begin{align}
    S_{\rm gen} =2S_{\rm gravity} +\frac{N}{6} \log \left(d^4_{12}e^{2\rho(I)}e^{2\rho(A)}\right)\,.
\end{align}
Recall that the gravity entropy is $S_{\rm gravity}(I)=\frac{N}{6}(\Om(I)-\rho(I))$ from Eq.~\eqref{Eq:2.last}, then the generalized entropy reads 
\begin{align}
     S_{\rm gen} \simeq & \frac{N}{3}(-x_I^+x_I^-+M)+\frac{N}{6}\log\left[(x_I^+-x_A^+)^2(x_A^--x_I^-)^2\right]\nonumber\\
     &+\frac{N}{6}\log\left( \frac{1}{-x_A^+x_A^-+M+\frac{a}{2}(-x_A^+x_A^-+M)}\right)\,,\label{eq:geneq}
\end{align}
where $x_I^\pm$ are the coordinates of the island boundary and $x^\pm_A$ are for the cutoff surface. Extremizing it with respect to $x_I^\pm$ gives 
\begin{align}
    \frac{\p S_{\rm gen}}{\p x_I^+}=-\frac{N}{3}x_I^-+\frac{N}{3(x_I^+-x_A^+)}=0\,,\\
     \frac{\p S_{\rm gen}}{\p x_I^-}=-\frac{N}{3}x_I^++\frac{N}{3(x_I^--x_A^-)}=0\,.
\end{align}
Taking the anchor point to be far from the horizon, $x_A^+\rightarrow \infty$, we have
\begin{align}
    x_I^\pm\approx -\frac{1}{x_A^\mp}\,,\ {\rm or}\quad \sigma^\pm_1=\sigma^\mp_2\,. 
\end{align}
Transforming to the $\{t,y\}$ coordinates, it gives the time condition of the island boundary $t_I=t_A$. The generalized entropy expressed in the coordinates $(t,y)$ reads
\begin{align}
    S\simeq\frac{N}{3}M+\frac{N}{3}\log\left[e^{2y}-e^{-2y}\right]+\frac{N}{6}\log\left(\frac{1}{M+e^{2y}+\frac{a}{2}\log(M+e^{2y})}\right) \,,
\end{align}
where $y$ is the spacial coordinate of point $A$. At late times, the entropy of the pair of black holes has an exponentially-suppressed dependence on the specific model parameterized by $``a"$, and is bounded by twice the Bekenstein-Hawking entropy instead of going to the infinity $ S\approx 2S_{\rm BH}$. The boundary points of the island are located approximately at $\sigma^+_I=\sigma^-_A$ at the horizon. The Page time $t_{\rm Page}\simeq M$ is approximately when the entropy without the island reaches the Bekenstein-Hawking value. It should be noticed that the island component is invariant under the change of theories in this parameter class as the indicated by Eq.~\eqref{eq:geneq}, the metric change due to the interpolation parameter only influences the CFT entropy. Therefore, the island configuration is the same for this class of solutions.

\subsection{Islands in dynamical black holes formed by collapsing shell}\label{subsub:bhshell}
We consider the case when the RST-BPP black holes is formed by collapsing matter shell. The shell of matter has the energy-momentum tensor $T_{++}= \frac{M}{x_0^+}\delta(x^+-x_0^+)$ in the Kruskal coordinates. In the regions inside the shell, the solution is required to be back to the vacuum solution. The geometry of the collapsing shell can be computed from Eq.~\eqref{Eq:eom} to be as follows,
\begin{align}
      \Om=-x^+x^--\frac{1}{4}\log(-x^+x^-)-M(x^+-x_0^+)\theta(x^+-x_0^+)\,,
\end{align}
up to a constant which can be determined by considering the endpoints of the evaporation. The apparent horizon ($\p_+\Om=0$) is located at $-x^+(x^-+M)=\frac{1}{4}$, and the vacuum energy is $t_\pm(x^\pm)=-\frac{1}{4(x^\pm)^2}$. Before the $x^+_0$, the (incoming) asymptotically flat coordinates are given by $\pm e^{\pm\sigma^\pm}=x^\pm$. After $x_0^+$, the (outgoing) asymptotically flat coordinates become 
\begin{align}
    x^+=e^{\t\sigma^+}\,, \quad -(x^-+M)=e^{-\t\sigma^-}\,,
\end{align}
and the metric is $ds^2\approx -d\t\sigma^+d\t\sigma^-$. In the outgoing asymptotically flat coordinates, the energy tensor of the radiation can be computed through the Schwarzian transformation \eqref{t_transform}
\begin{align}
    t_+(\t\sigma^+)=0\,,\quad t_-(\t\sigma^-)=\frac{1}{4}-\frac{(x^-+M)^2}{4(x^-)^2}\,.
\end{align}

In the late times $\t\sigma^-\rightarrow\infty$, the coordinate-dependent terms in $t_-(\t\sigma^-)$ vanishes as $x^-+M\rightarrow 0$, the vacuum energy flux only has a constant outgoing component $t_-(\t\sigma^-)=\frac{1}{4}$, which corresponds to a constant temperature $\frac{1}{2\pi}$. Therefore, the black hole has an evaporation time which is approximately $4M$. This time can be calculated exactly considering the dynamics of the black hole. The evaporation process ends when the apparent horizon hits the singularity at $x^-_{\rm int}=\frac{-M}{1-e^{-4M+2}}$. This corresponds to the evaporation time $\t\sigma^-_{\rm int}=4M-\log M$, which is consistent with the approximation from the energy current consideration. The geometry after the apparent horizon hits the singularity can be continuously mapped to a dilaton vacuum through a thunderbolt emission, an instantaneous emanation of some finite amount of energy in the form of radiation. For the BPP model $a=0$, the resulting vacuum geometry has an infinite throat and is geodesic complete, otherwise it is the vacuum solution with a boundary. However, the details near this endpoint and the vacuum geometry will not influence the overall analysis in this paper. 

\begin{figure}[tb]
\centering
	\includegraphics[width=70mm]{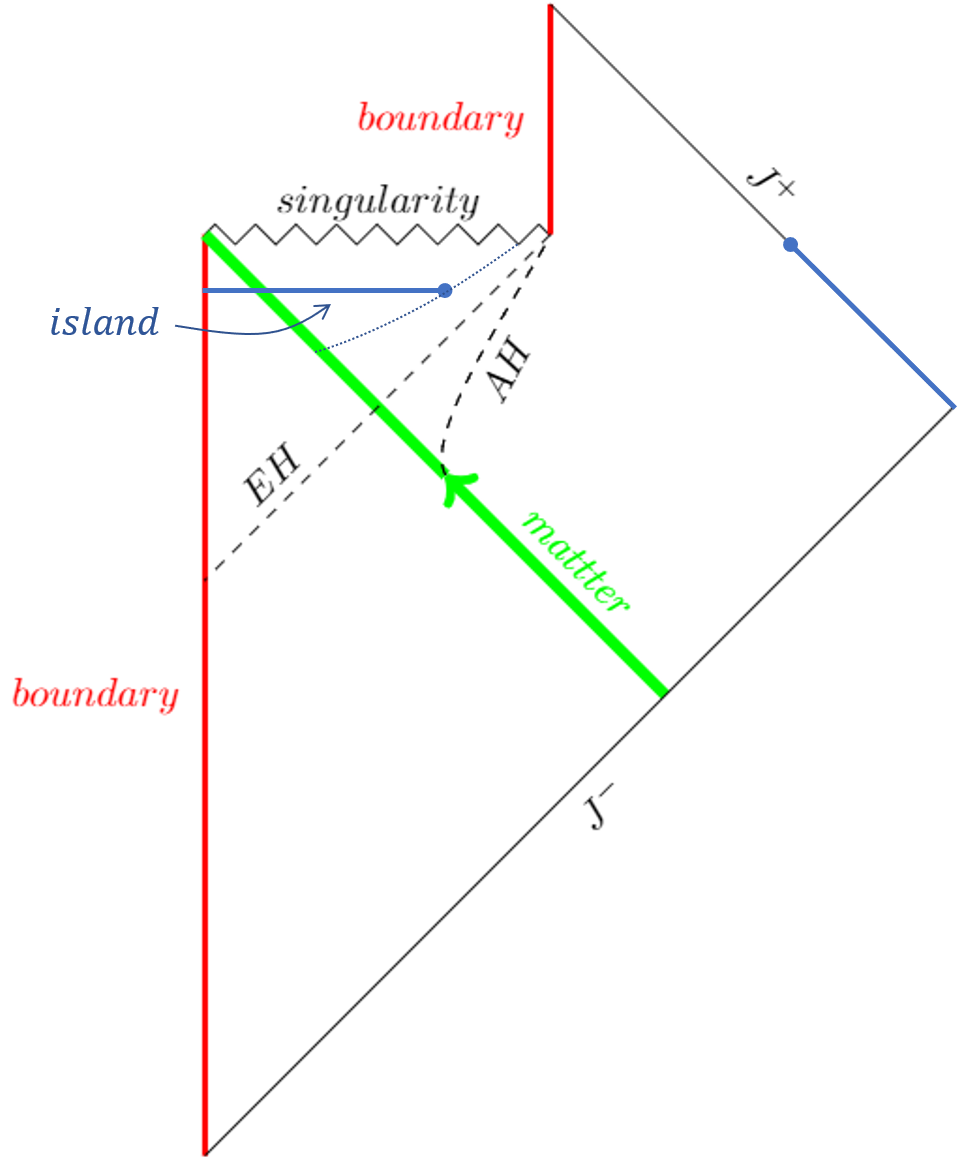}
	\caption{The Penrose diagram for the formation of a dilaton black hole by a collapsing shell of an f wave. The vertical red lines are the boundaries of the spacetime. The green line represents the infalling matter wave. EH stands for the event horizon and AH is the apparent horizon. The radiation is collected near the null infinity as indicated by the blue line along $J^+$.}
	\label{fig:eva_island}
\end{figure}

The entanglement entropy for a 2-dimensional vacuum CFT bounded by one endpoint is given by \cite{Fiola:1994ir}, 
\begin{align}
    S_{\rm ent}=\frac{N}{6}\left[\rho(\sigma^+,\sigma^-)+\log\frac{\sigma^+-\sigma^-}{\epsilon_{uv}}\right]\,,\label{Eq:s_cft}
\end{align}
where $\sigma^\pm$ are the incoming asymptotically flat coordinates of the cutoff point, $\epsilon_{uv}$ is the UV cutoff and the conformal factor $\rho(x^+,x^-)$ can be calculated from
\begin{align}
      \Om_a=e^{-2\rho}+a\rho=-x^+x^--\frac{1}{4}\log(-x^+x^-)-M(x^+-x_0^+)\theta(x^+-x_0^+)\,.
\end{align}
It can be easily shown that $\rho(\sigma^+,\sigma^-)=\rho(x^+,x^-)+\sigma$, where $\sigma=\frac{1}{2}(\sigma^+-\sigma^-)$. For the cutoff point near the future null infinity $J^+$, the divergent terms becomes independent of the retarded time and can be properly ignored. The entanglement entropy reads
\begin{align}
    S=\frac{N}{6}\rho_a(\sigma^+,\sigma^-)=\frac{N}{12}\left(2\rho_a(x^+,x^-)+\sigma^+-\sigma^-\right)\,,
\end{align}
where the second term in Eq.~\eqref{Eq:s_cft} is ignored since it is the time-independent divergent part of the CFT entropy and is not related to the time-dependent entanglement entropy of the Hawking radiation. The conformal factor can be approximately expressed as
\begin{align}
    2\rho_a(x^+,x^-)\approx& -\log\left[-x^+x^--\frac{1}{4}\log(-x^+x^-)-M(x^+-x_0^+)\right.\nonumber\\
    &\qquad\quad +\left.\frac{a}{2}\log(-x^+x^--\frac{1}{4}\log(-x^+x^-)-M(x^+-x_0^+))\right]\,.
\end{align}
Pushing the cutoff point to the conformal boundary $J^+$, the conformal factor to the leading order is independent of the parameter ``$a$" and approximately reads
\begin{align}
    2\rho_a(x^+,x^-)\approx & -(\sigma^+-\sigma^-)-\log\left(1+\frac{M}{x^-}\right)\nonumber\\
    \approx&\log\left(1+Me^{\t\sigma^-}\right)-2\sigma\,.
\end{align}
Therefore, at the late times the entropy at the zeroth order is independent of parameter $a$ and increases linearly with the retarded time in a simple manner $S\sim \frac{N}{12}\t\sigma^-$,
where $\t\sigma^-$ is the affine time at the future null infinity. 

The calculation of entanglement entropy of the black hole without island can also be done holographically. If we pick one of the two endpoints near the singularity, the semi-classical method breaks down near that region. As is argued in \cite{Gautason:2020tmk} for the RST model, under the assumption that this endpoint near the singularity does not change dramatically with the time, and then we can proceed by factoring out the time-independent piece and computing the entropy change which then only depends upon the asymptotic time on the cutoff surface. This approach, though less rigorous, gives the same result as we presented above for the RST model.  

The entanglement entropy of the black hole considering the configuration of the island can be calculated holographically assuming that the boundary of the island is far away from the singular points. We will see that this assumption is valid until the end of the evaporation process when the apparent horizon hits the singularity. This strongly coupled regime only appears in the end process and the details of it does not affect the validity of our analysis which applies in a much larger time scale of the black hole lifetime. 

The entanglement entropy of the black hole is the minimum of the extremal values of the sum of contributions---the gravity part plus the bulk entanglement,
\begin{align}
     S_{\rm gen}(I\cup R)=S_{\rm gravity}+S_{\rm bulk}\,.
\end{align}
The holographic entanglement of the bulk is
\begin{align}
    S_{\rm bulk}= \left. \frac{N}{6} \log \left[d(I,A)^2e^{\rho(I)}e^{\rho(A)}\right]\right|_{t_\pm=0} \, ,
\end{align}
where $d(I,A)$ is the two-dimensional distance computed in the flat coordinates obtained by a Weyl transformation of the original metric. $I$ and $A$ are the boundary points of the island and the cutoff surface, respectively. 

\begin{figure}[t]
\centering
	\includegraphics[width=100mm]{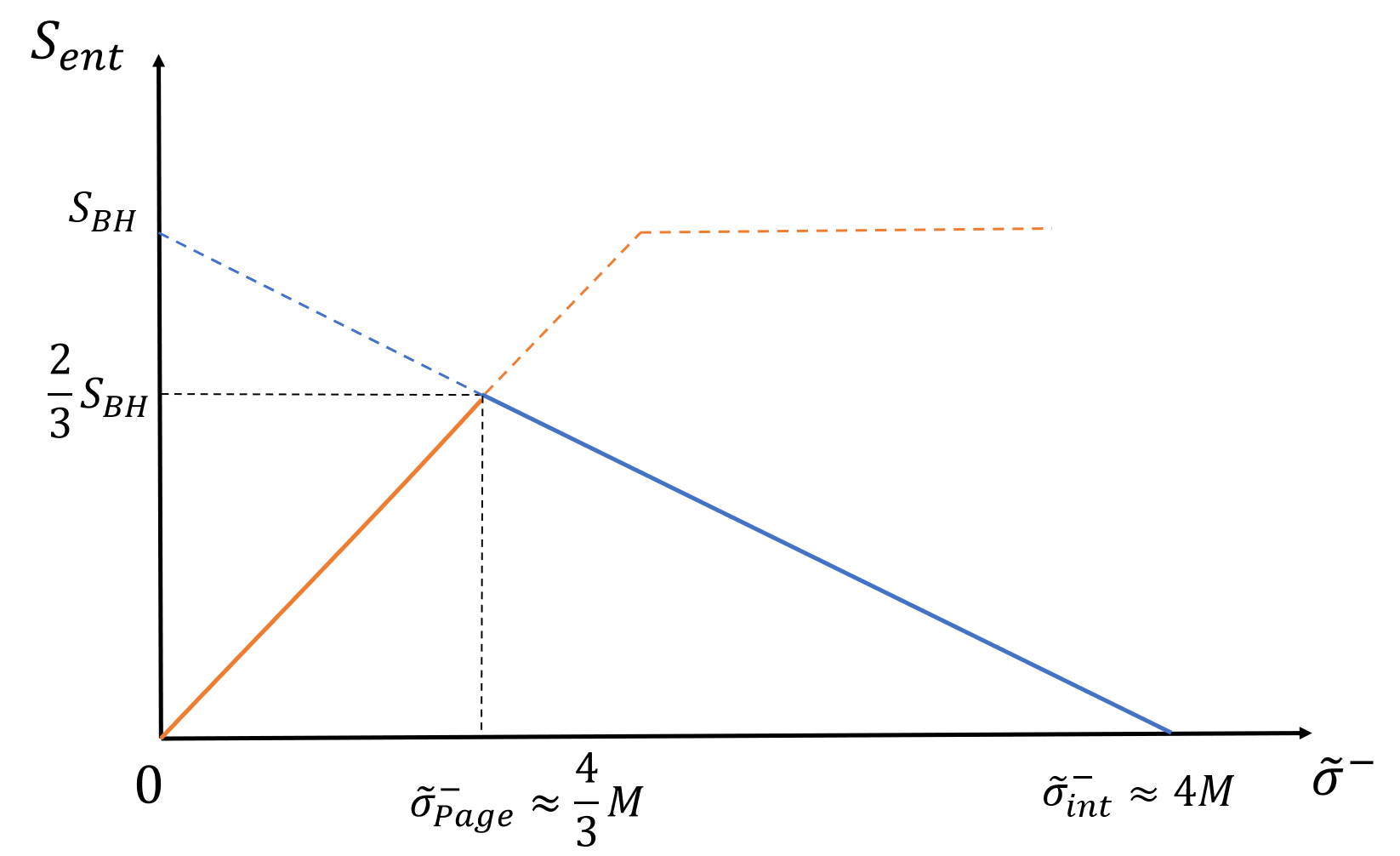}
	\caption{The Page curve for an evaporating black hole formed by a shell collapse.}
	\label{fig:page_evaporating}
\end{figure}

The ingoing and outgoing waves in the Kruskal frame are in a state with constant temperature $\frac{1}{2\pi}$. These modes can be mapped to the vacuum CFT state by the logarithmic mapping, which can be obtained by solving the anomalous transformation equation, 
\begin{align}
 0=t_{\pm}(x^\pm)+\frac{1}{2}\{\sigma^\pm,x^\pm\}\,.
\end{align}
The solution of the above equations is the incoming asymptotically flat coordinates $\{\sigma^+,\sigma^-\}$ defined by $e^{\pm\sigma^\pm}=\pm x^\pm$. Pushing the anchor point $A$ to the future null infinity and ignoring the UV and IR cutoff terms, the generalized entropy is 
\begin{equation}
    S_{\rm gen}(I\cup R)=S_{\rm gravity}+S_{\rm bulk}=\frac{N}{6}\left[\Omega(I)+\frac{1}{2}(\sigma^+_I-\sigma^-_I)+\log|\sigma^-_I-\sigma^-_A|\right] \,.
\end{equation}
Extremize it with respect to $\sigma^\pm_I$, we have
\begin{align}
    S_{\rm gen}(I\cup R)\approx \frac{N}{6}M-\frac{N}{24}\tilde\sigma^-_A\,,
\end{align}
and the location of the island satisfies $x^+_I(x^-_I+M)=\frac{1}{4}$, which lies inside the event horizon. The first term is the Bekenstein-Hawking entropy, and the second term shows that the entropy of the black hole peaks at $\frac{1}{3}$ of its evaporation time. The maximal value of the fine-grained entropy is $S_{\rm max}=\frac{2}{3}S_{\rm BH}$. The black hole entropy decreases to zero at the retarded time $\t\sigma^-_A=4M$, which is the lifetime of the evaporating black hole shown in Sec.~\ref{subsub:bhshell}. Omitting the subscript ``A", the entanglement entropy is 
\begin{align}
    S_{\rm ent}=\min(\frac{N}{12}\t\sigma^-_,S_{\rm BH}-\frac{N}{24}\t\sigma^-)\,,
\end{align}
which grows at the rate $\frac{N}{12}\t\sigma^-$ and decays at $\frac{N}{24}\t\sigma^-$ as shown in Fig.~\ref{fig:page_evaporating}. A similar result was also derived in the RST model \cite{Fiola:1994ir,Gautason:2020tmk}. The island collides with the singularity when the observer time is $\t\sigma^-_{\rm col}=4M-\log M-\log 3 -2$, which is approximately the time when the black hole evaporates completely in the large-mass limit. Therefore, this analysis can be trusted all the way to the final stage of the evaporation. This result in the leading order approximation is independent of the specific models parameterized by $``a"$. Therefore, we can conclude that the islands and Page curves in the leading order are the same for this variety of 2D asymptotically-flat gravities.

\section{Gluing equilibrium black holes with evaporating ones}
It is beneficial to extend the studies of the entanglement entropy of the radiation beyond the two simple scenarios and consider the effect of gluing two different geometries. In this section, we consider the case when the black hole is initially in the Hartle-Hawking state. Then we cut the Penrose diagram along a null line and glue the evaporating black hole geometry along the null trajectory. We discuss the effect of gluing and corresponding island configurations in the lowest order. In the previous section, we have shown that the free parameter $``a"$ interpolating between different theories contribute only to the higher order of the entanglement entropy at the large distances and limit times and does not alter the qualitative picture of the Page curve. Thereofre, in this section we will content ourselves with the leading order results.

\subsection{Geometry}
\begin{figure}[tbh]
\centering
	\includegraphics[width=100mm]{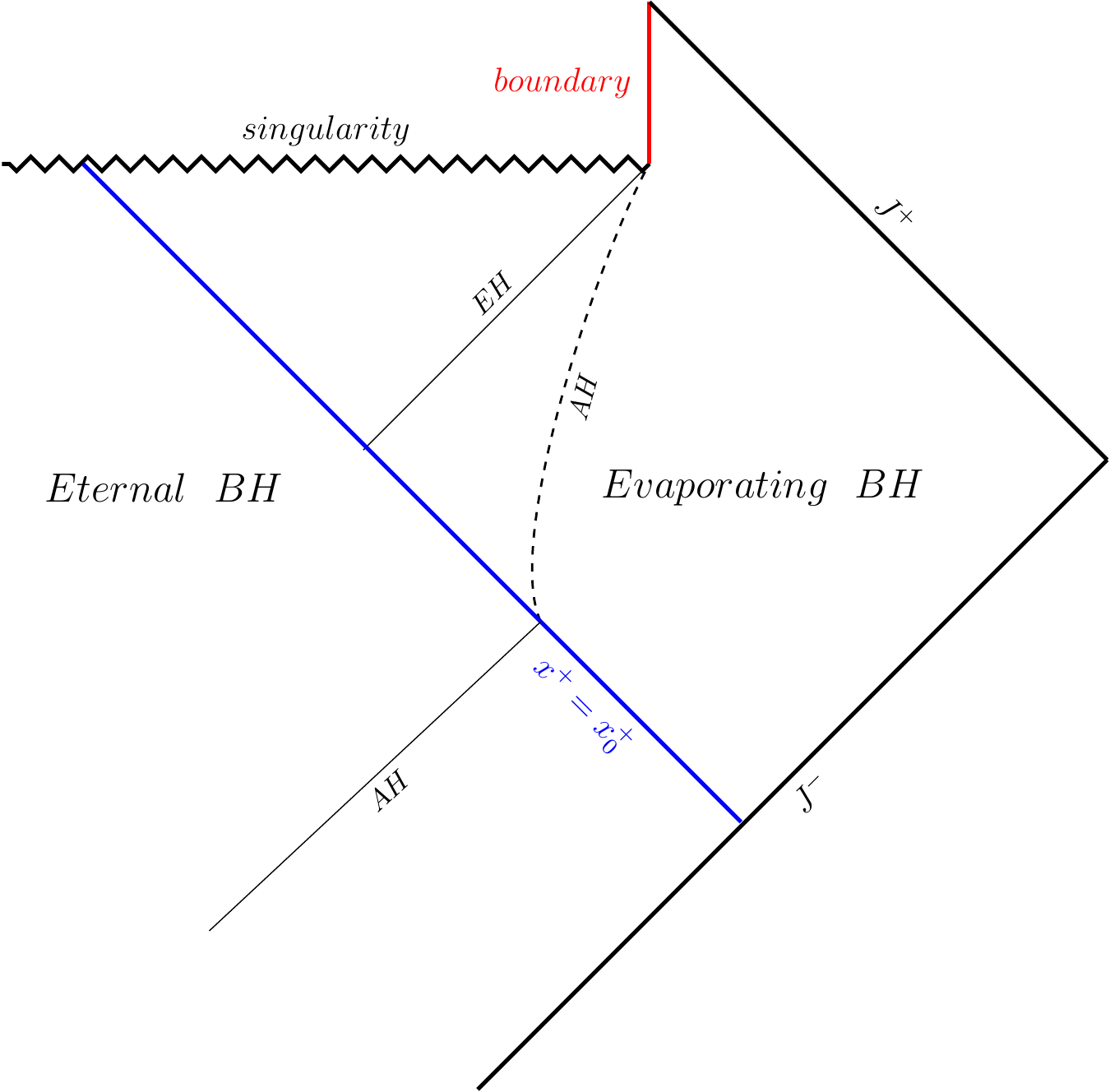}
	\caption{Sketch of the Penrose diagram for an black hole with thermal bath removed at $x_0^+$. To the left of the null line $x_0^+$, the geometry is identical to that of an eternal black hole or the black hole in equilibrium with the infalling radiation emitted from an external source. ``AH" represents the apparent horizon, ``EH" stands for the event horizon. $J^-$ and $J^+$ represent the past and future null infinity, respectively. After the evaporation, the space returns to the asymptotically flat space with the boundary.}
	\label{fig:Penrose_removal}
\end{figure}

We assume that the black hole has been in equilibrium with the bath until we remove the bath, and the black hole starts to evaporate henceforth. Therefore, in the asymptotically flat coordinates $\sigma^\pm$, there is no incoming radiation after the null line $x^+>x^+_0$ and the outgoing radiation $t_-(\sigma^-)=\frac{1}{4}$ is the same as that of the initial black hole in equilibrium. Transforming back to the $x^\pm$ coordinates, the corresponding boundary conditions are \footnote{Note that we have a sign difference from \cite{Cruz:1996pg} due to different conventions in the equations of motion.}
\begin{gather}
    t_+(x^+)=-\frac{1}{4(x^+)^2}\theta(x^+-x_0^+)\,,\\
    t_-(x^-)=0\,.
\end{gather}
Putting the boundary conditions back to Eq.~\eqref{Eq:Omega}, we have for $x^+>x_0^+$,
\beq
\Om=-x^+(x^-+\Delta)-\frac{1}{4}\left(\log\frac{x^+}{x^+_0}+1\right)+M\,,
\eeq
where $\Delta=-\frac{1}{4x^+_0}$. For $x^+<x_0^+$, the solution for $\Omega$ is the eternal black hole solution. The apparent horizon at $x_0^+$ satisfies the equation $\left.\p_+\Om\right|_{x^+_0} =-x^-=0$ and is glued continuously at $x^+=x_0^+$ with the apparent horizon of the eternal black hole. A discontinuity appears crossing $x_0^+$ from the horizon of the equilibrium black hole geometry to the event horizon of the bath-removed black hole geometry. The event horizon lies inside the apparent horizon for the evaporating black hole as shown in Fig.~\ref{fig:Penrose_removal}. The evaporation process ends when the apparent horizon reaches the singularity, which occurs at
\begin{align}
    x_{\rm int}^-=\frac{1}{4x_0^+}(1-e^{-4(M-1/4)})\quad {\rm or}\quad \t\sigma^-_{\rm int}= 4M+\log(4x^+_0)-1\,,
\end{align}
where coordinates $\{\t\sigma^+,\t\sigma^-\}$ are now the outgoing asymptotically flat coordinates of the bath-removed geometry defined by 
\begin{align}
    x^+=e^{\tilde\sigma^+}\,,\quad x^-+\Delta=-e^{-\tilde\sigma^-}\,.\label{co:out}
\end{align}
Similarly, it can be shown that the endstate of the black hole and the resulting geometry are the same as that in the evaporating black hole case.

\subsection{Islands in bath-removed space}
For simplicity, we first assume that the thermal bath is removed at a very early stage $x^+=x^+_0\ll M$ and when the black hole is in the pure state. Therefore, we can ignore the geometry before $x^+_0$ and only analyze the geometry after $x^+_0$. 

The appropriate coordinates $U^\pm$ for computing the vacuum entanglement of CFT, which solves the equation of the anomalous transformation $ 0=t_{\pm}(x^\pm)+\frac{1}{2}\{U^\pm,x^\pm\}$ is $(\sigma^+,\ x^-)$. For the dynamical solution after $x^+_0$, 
\beq
\Om_a=e^{-2\rho}+a\rho=-x^+(x^-+\Delta)-\frac{1}{4}\left(\log\frac{x^+}{x^+_0}+1\right)+M\,,
\eeq
where $\Delta=-\frac{1}{4x_0}$ and $x_0$ is the null line along which the bath is removed. One can check that the geometry connects continuously with the eternal black hole geometry $\Om_{\rm eternal}$ at $x_0^+$. At the end of the evaporation it connects continuously with a vacuum dilaton solution of which the detail does not concern us here. 

Ignoring the divergent terms (UV and IR cutoffs) which are independent of the asymptotic time, the entanglement entropy of the CFT is
\begin{align}
    S&=\frac{N}{6}\rho(\t\sigma^+,x^-)=\frac{N}{12}(2\rho(x^+,x^-)+\t\sigma^+)\nonumber\\
    &\approx -\frac{N}{12}\log\left[-(x^-+\Delta)+e^{-\tilde\sigma^+}(\bar{M}-\tilde\sigma^+/4)\nonumber\right]\\
    &\approx \frac{N}{12}\tilde\sigma^-\,,
\end{align}
where $\t\sigma^\pm$ are the outgoing asymptotically flat coordinates defined in Eq.~\eqref{co:out} and 
\begin{align}
    \bar{M}=M+\frac{1}{4}\log x^+_0-\frac{1}{4}\,.
\end{align} 
The entropy grows linearly with respect to the affine time which apparently violates the unitarity at the late times. The fine-grained entropy of the black hole is bounded by the Bekenstein-Hawking entropy regardless of the existence of the external bath. 


We consider the situation that allows the appearance of an island. We assume that the cutoff surface is around the future null infinity such that we can approximate the conformal factor as 
\begin{align}
    e^{-2\rho(\sigma^\pm)}=1\,
\end{align}
along the world line of the observer. The generalized entropy with islands can be shown to take the following form,
\begin{equation}
    S_{\rm gen}(I\cup R)=S_{\rm gravity}+S_{\rm bulk}=\frac{N}{6}\left[\Omega(I)+\frac{1}{2}\sigma^+_I+\log|x^-_I-x^-_A|+\frac{1}{2}\t\sigma^-_A\right] \,.\label{eq:gen_island_bath}
\end{equation}
Extremizing the generalized entropy with respect to the boundary coordinates of the island $\sigma^+_I$ and $x^-_I$ gives, 
\beq
e^{\sigma^+_I}=-\frac{3}{4(x^-_A+\Delta)}\,, \quad x^-_I=-\frac{x^-_A+4\Delta}{3}\,. \label{Eq:removal_island_position}
\eeq
The position of the island satisfies 
\begin{align}
     x^+_I(x^-_I+\Delta)=\frac{1}{4}\,,
\end{align}
which lies inside the apparent horizon at $x^+(x^-+\Delta)=-\frac{1}{4}$ and the event horizon which locates approximately at $x^+(x^-+\Delta)=0$.

\begin{figure}[tb]
\centering
	\includegraphics[width=110mm]{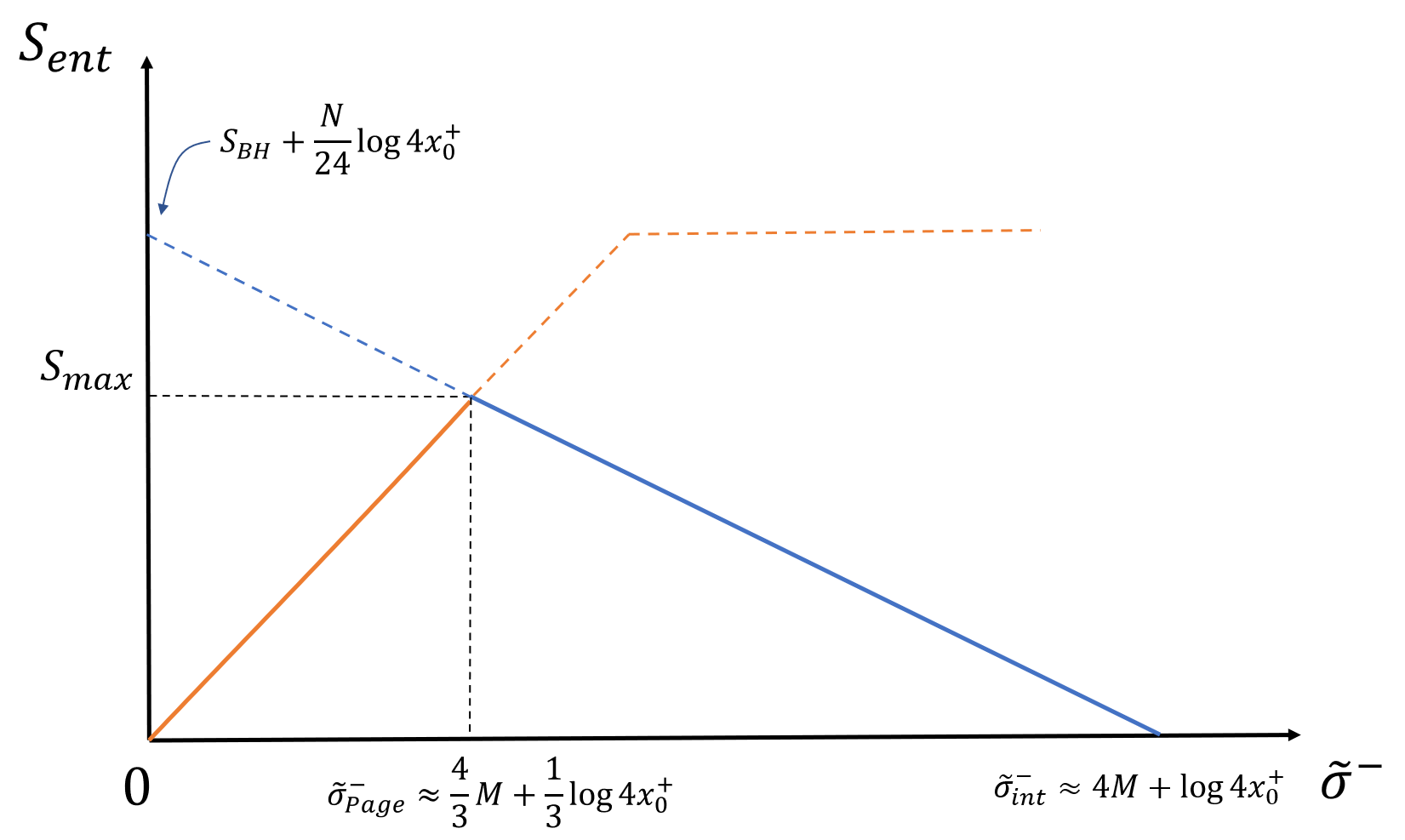}
	\caption{The Page curve for a bath-removed evaporating black hole.}
	\label{fig:page_removal}
\end{figure}

Knowing the relation between the coordinates of the island and those of the observer, we have the quantum surface area $\Om$ of the island given as follows
\begin{align}
    \Omega(I)&=-(-\frac{3}{4(x^-_A+\Delta)})(-\frac{x^-_A+4\Delta}{3}+\Delta)-\frac{1}{4}\left(\log( -\frac{3}{4(x^-_A+\Delta)x_0^+})+1\right)+M\nonumber\\
    &=-1/4-\frac{1}{4}\left(\log( -\frac{3}{4(x^-_A+\Delta)x_0^+})+1\right)+M\,.
\end{align}
The generalized entropy with the island in Eq.~\eqref{eq:gen_island_bath} is
\begin{align}
    S_{\rm gen}(I\cup R) &=\frac{N}{6}\left[\Omega(I)+\frac{1}{2}\log(-\frac{3}{4(x^-_A+\Delta)})+\log(-\frac{4x^-_A+4\Delta}{3})+\frac{1}{2}\t\sigma^-_A\right]\nonumber\\
    &=\frac{N}{6}\left[-1/4-\frac{1}{4}\left(\log( \frac{3}{4e^{-\tilde\sigma_A^-}x_0^+})+1\right)+M+\frac{1}{2}\log(-\frac{4x^-_A+4\Delta}{3})+\frac{1}{2}\t\sigma^-_A\right]
\end{align}
Using the asymptotically flat coordinates $\{\tilde\sigma^+,\tilde\sigma^-\}$, we have the generalized entropy with island computed as follows,
\begin{align}
    S_{\rm gen}(I\cup R)&=\frac{N}{6}\left[-1/4-\frac{1}{4}\left(\log( \frac{3}{4e^{-\tilde\sigma_A^-}x_0^+})+1\right)+M+\frac{1}{2}\log(\frac{4e^{-\tilde\sigma_A^-}}{3})+\frac{1}{2}\t\sigma^-_A\right]\nonumber\\
    &\approx \frac{N}{6}M-\frac{N}{24}\tilde\sigma_A^-+\frac{N}{24}\log(4x_0^+)\\
    &\approx \frac{N}{24}(\t\sigma^-_{\rm int} - \tilde\sigma_A^-)
    \,.\label{Eq:removalisland}
    \end{align}
Note that the first term $\frac{N}{6}M=S_{\rm BH}$ is the Bekenstein-Hawking entropy of a single black hole of mass $M$. The generalized entropy decreases linearly with the affine time in the manner of $-\frac{N}{24}\t\sigma^-_A$. The fine-grained entropy of the black hole is the minimal generalized entropy of all configurations of the islands, in this case, the minimum of the $S_{\rm no-island}$ and $S_{\rm island}$. Therefore, we have the result of the fine-grained entropy \footnote{The subscript ``A" is omitted since the fine-grained entropy cannot explicitly depend on the island coordinates and is only dependent on the retarded time of the observer at $A$.} 
\beq S_{\rm ent}=\min\left(\frac{N}{12}\t\sigma^-\ ,\ \frac{N}{24}(\t\sigma^-_{\rm int} - \tilde\sigma^-)\right)\,,
\eeq
which follows the Page curve shown in Fig.~\ref{fig:page_removal}. The Page time appears at a third of the black hole evaporating time.


\subsection{Islands in general glued space}
If we relax the constraint that $x_0^+\ll M$, then we need to consider both the early radiation released before $x^+_0$ and after that. Before $\t\sigma_0^-$, the radiations reaching the cutoff surface are released when the black hole has a constant mass due to the incoming energy flux. Therefore, the island configuration and the Page curve should be referred to the eternal black hole case. For a one-sided black hole, we assume that the incoming energy flux is in a pure state to avoid counting the incoming radiation entropy and the complication of the discontinuity when the bath is removed.

\begin{figure}[tbh]
\centering
	\includegraphics[width=110mm]{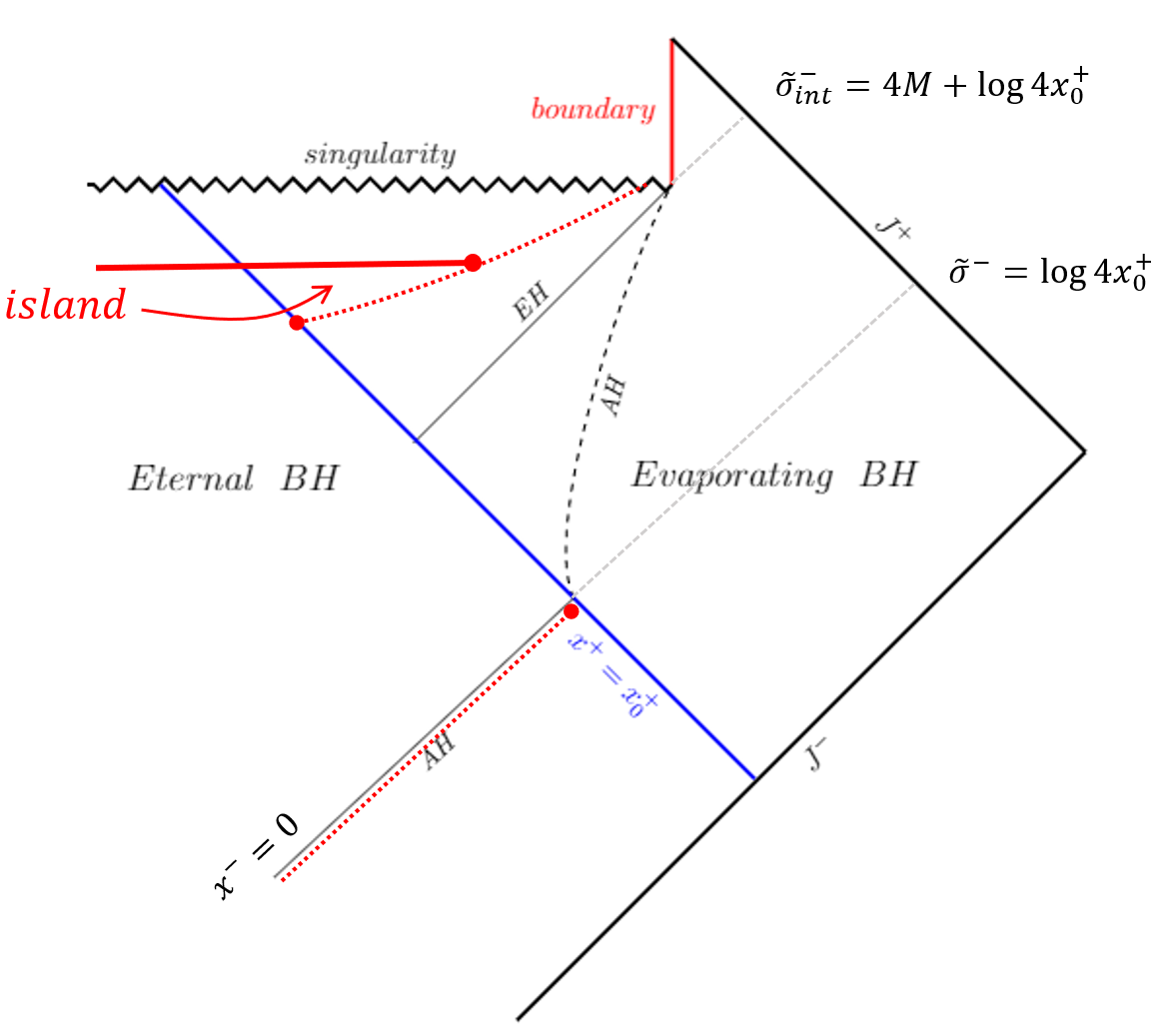}
	\caption{Sketch of the Penrose diagram for an black hole with thermal bath removed at $x_0^+$. The island is colored in red, and the trajectory of the island boundary is shown in red dotted lines. The $x^+_0$ null ray intersects with the apparent horizon at the retarded time $\t\sigma^-=\log 4 x_0^+$.}
	\label{fig:Penrose_removal_gen}
\end{figure}

We define the ``last eternal radiation surface" as the shell of the outgoing radiations which is emanated when the last beam of incoming radiation reaches the apparent horizon. Before the last eternal radiation surface hits the cutoff surface, namely $\t\sigma^-<\log 4x^+_0$, the incoming asymptotic time is approximately equal to the outgoing asymptotic time $\t\sigma^-\approx \sigma^-$. Before this point, the radiations at the null infinity are from the black hole in the equilibrium state and the entanglement entropy of the black hole is identical to that given in the eternal cases. Expressed in the $\{\t\sigma^+,\t\sigma^-\}$ coordinates, the renormalized retarded-time-dependent component of the generalized entropy without considering the island is
\begin{align}
    S=\frac{N}{12}\sigma^-\approx\frac{N}{12}\t\sigma^-\,.
\end{align}
Note that this is the same formula of the eternal black hole up to a coordinate transformation and keeping only the finite terms. It can also be derived separately using the method shown in the previous section. The island configuration before the $\t\sigma^-=\log 4x^+_0$ is the same as the eternal black hole scenario with only right half of the Penrose diagram. For $\log 4x^+_0>\t\sigma^-_{\rm Page}=2M$, the island appears approximately on the surface of the horizon before the last incoming radiation reaches the horizon. The boundary of the island will be carried by the last incoming null ray to the inside of the event horizon at $x_0^+$ satisfying the equation $x^+_I(x^-_I+\Delta)=\frac{1}{4}$. As shown in Fig.~\ref{fig:Penrose_removal_gen}, the boundary of the island jumps from the surface of the apparent horizon $x^-=0$ to the inside of the event horizon at $x_I^-=\frac{1}{2x_0^+}$ according to Eq.~\eqref{Eq:removal_island_position}.

In order for the island to exist we also need to require that the position of the island lies outside the singular boundary $\Om>\frac{1}{4}$ and beyond $x_0^+$. The island intercepts with the boundary at 
\begin{align}    x^+_{\rm int}=x^+_0\cdot e^{4M-3},  \quad x^-_{\rm int}=\frac{1}{4x_0^+}(1+e^{3-4M})\,, \end{align} 
which gives the retarded time of the observer in the outgoing asymptotically flat coordinates
\begin{align}
    \t\sigma^-_{\rm int}= 4M+\log(4x_0^+)-\log 3-3\,.
\end{align}
Note that this is approximately (slightly before) the time when the apparent horizon hits the singularity and the evaporation process ends. Since the difference is negligible (of order $M^{-1}$), we have used the same notation for the intersecting times. This difference does not influence our analysis. The condition for the endpoint of the island to be outside $x_0^+$ gives 
\begin{align}
    \t\sigma^->\log 4x^+_0-\log3\,,
\end{align}
which is the condition for the anchor points to be chosen approximately after the infalling null ray at $x_0^+$ crosses the horizon or slightly before this time. This equation shows that the island configuration can be trusted after the ``last eternal radiation" reaches the cutoff surface. Therefore, the entanglement entropy is given by 
\begin{align}
    S\simeq \left\{
    \begin{matrix} 
        &\min (\frac{N}{12}\t\sigma^-,\frac{N}{6}M)\,, \ {\rm when}\ \t\sigma^-<\log 4x^+_0\,, \\ &\min\left(\frac{N}{12}\t\sigma^-,\frac{N}{6}M-\frac{N}{24}(\t\sigma^--\log 4x^+_0)\right)\,,\ {\rm when}\ \t\sigma^->\log 4x^+_0\,.
    \end{matrix}\right.
\end{align}

\begin{figure}[htb]
\centering
	\includegraphics[width=110mm]{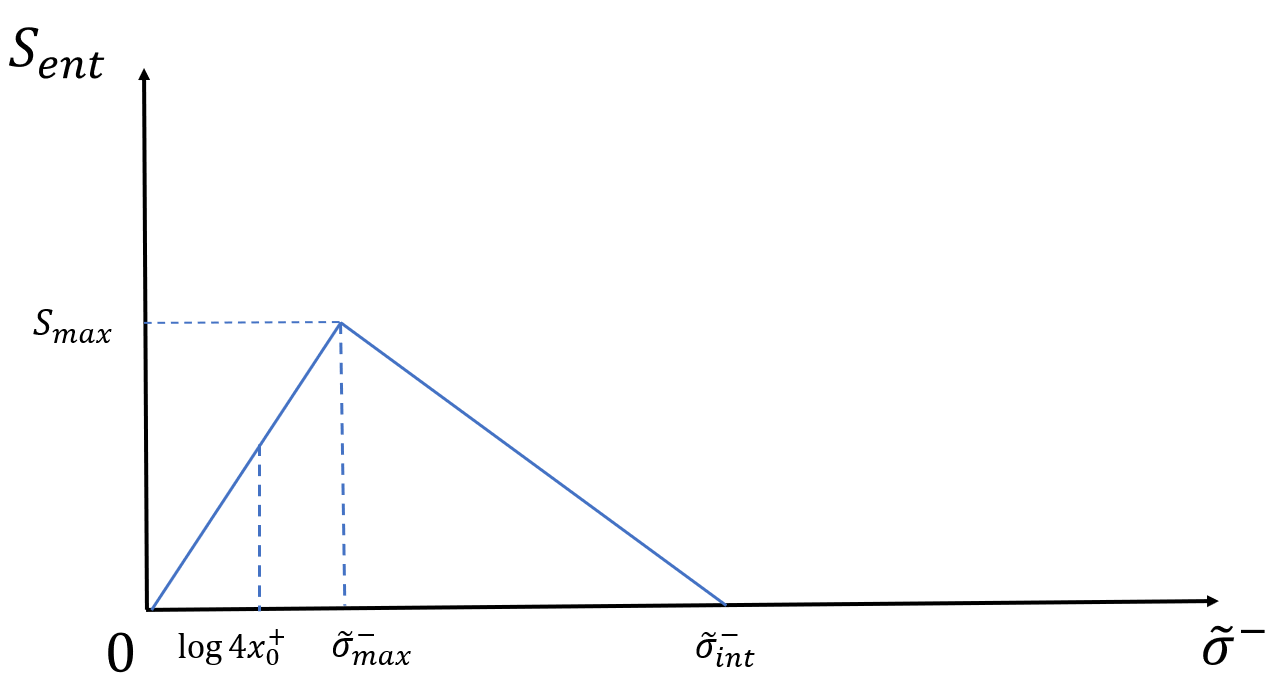}
	\includegraphics[width=110mm]{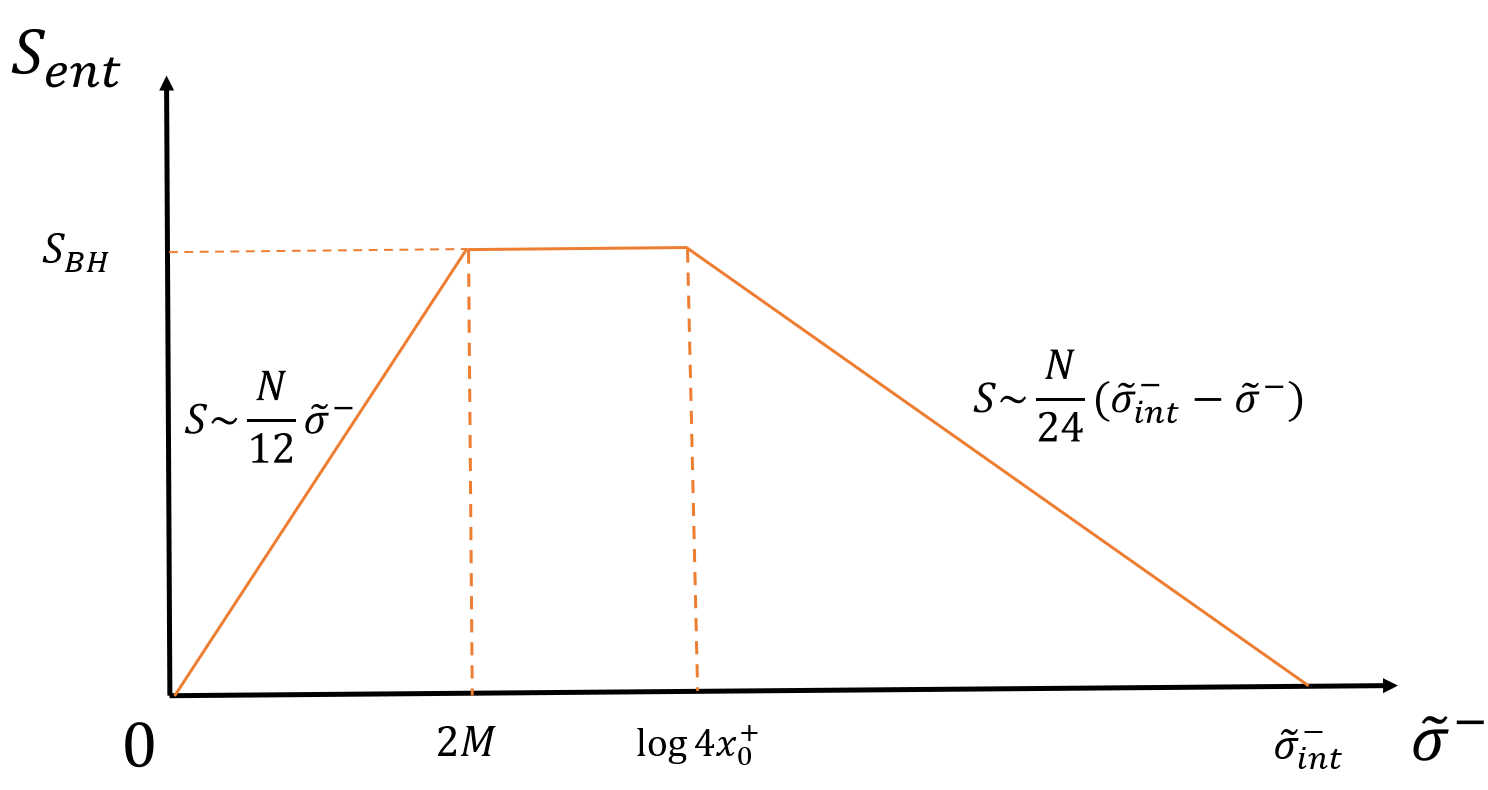}
	\caption{Sketch of the Page curves with the thermal bath removed along the null trajectory $x^+=x_0^+$. (a) The blue curve represents the case of early bath removal $x^+_0<\frac{e^{2M}}{4}$, where the maximal entanglement entropy is $S_{\rm max}=\frac{2}{3}(S_{\rm BH}+\frac{N}{24}\log4x^+_0)$ at the time $\t\sigma^-_{\rm max}=\frac{4}{3}M+\frac{1}{3}\log 4x^+_0$. (b) The orange curve is the Page curve for the late bath removal scenario, $x^+_0>\frac{e^{2M}}{4}$. The first transition time (Page time) is around the retarded time $\t\sigma^-=2M$ when the island contributing to the entanglement entropy appears near the apparent horizon of the equilibrium black hole. The second transition time shows up at $\t\sigma^-=\log 4x^+_0$ when the boundary of the island jumps to inside the event horizon. The endpoint of evaporation time $\t\sigma_{int}^-\simeq 4M+\log(4x_0^+)$ is $x_0^+$ dependent and is larger for the second case. }
	\label{fig:page_gen_removal}
\end{figure}

Though the island changes discontinuously at the time $\tilde\sigma=\log 4x^+_0$, the entanglement entropy is continuous. Since the generalized entropy is the sum of the island area and the bulk entropy, the shrinking of island areas suggests an increase of entropy outside the island. This means that a large amount of entropy is stored near the horizon and these quanta are mixed state entangled with the radiation. For the case of an early bath removal $\log 4x^+_0<2M$, the island appears only in the bath-removed sector of the spacetime. The entanglement entropy in late times decays roughly in the manner of $\frac{N}{24}(\t\sigma^-_{\rm int}-\t\sigma^-)$.  The maximal entropy is reached when $\frac{N}{12}\t\sigma^-=\frac{N}{6}M-\frac{N}{24}(\t\sigma^--\log 4x^+_0)$. This gives the maximal entanglement entropy 
\begin{align}
    S_{\rm max}=\frac{2}{3}(S_{\rm BH}+\frac{N}{24}\log4x^+_0)<S_{\rm BH}\,,
\end{align} 
which appears at the retarded time $\t\sigma^-_{\rm max}=\frac{4}{3}M+\frac{1}{3}\log 4x^+_0$ [blue curve in Fig.~\ref{fig:page_gen_removal}].

For the case of a late bath removal $\log 4x^+_0<2M$, the island appears both in the eternal black hole region and the bath-removed region. The entanglement entropy follows that of an eternal black hole scenario at the beginning of the evaporation process which increases linearly  until the fine-grained entropy reaches values of the coarse-grained entropy. Due to the appearance of the island in the equilibrium black hole geometry, the fine-grained entropy stays at the Bekenstein-Hawking value before the retarded time $\t\sigma^-=\log 4x^+_0$. The boundary position of the island experienced a jump at the retarded time $\t\sigma^-=\log 4x^+_0$ from slightly outside the apparent horizon of the ``eternal" black hole to inside the event horizon as shown in Fig.~\ref{fig:Penrose_removal_gen}. The entropy then follows a different time curve given by $\frac{N}{24}(\t\sigma^-_{\rm int}-\t\sigma^-)$. The Page curve is shown in Fig.~\ref{fig:page_gen_removal}.

The Page times in Figs.~\ref{fig:page_gen_removal} and \ref{fig:page_removal}(a) that appear at a third of the evaporating process can be interpreted following a simple thermodynamic argument. In the initially stage, the fine-grained entropy of the radiation is approximately its thermal entropy, which is $S_{gas}\simeq 2U/T= -2\Delta M/T$, where $\Delta M$ is the change of black hole mass. On the other hand, we can also draw the Bekenstein-Hawking entropy of the black hole $S_{\rm BH}=M/T$. Since the fine-grained entropy of the radiation is equal to that of the black hole, these two lines join at the point when black hole has evaporated a third of its mass and the Page curve is simply the minimum of the two curves. From this argument, it is natural that the entropy drops at half of the rate it rises in the initially period. Though this reasoning is simple and is supposed to be valid at the initial and final states of evaporation assuming that the radiation and the black hole states are randomly chosen in uniform way from the total states with respect to a Haar measure, it has some issues when we think beyond the thermodynamic argument. In the line of this reasoning, we have assumed that the radiation in the initial state is highly entangled with the black hole, and in the final state the black hole is highly entangled with the early radiation \footnote{This is true if we average over all states in the product Hilbert space of the black hole and the radiation $\mathcal{H}_{BH}\otimes \mathcal{H}_R$. Modifications can be applied when considering only the states with a fixed total energy, and this results in the thermal density matrix in the reduced system. However, the difference does not qualitatively change the result of the analysis.}. However, according to Page's theorem (which assumes that the black hole can be described as an ordinary quantum system), these two processes are equivalent up to a time reversal given that the 2D black hole radiates at a constant rate and temperature \cite{Page:1993wv}. Therefore, instead of looking at the black hole Bekenstein-Hawking entropy drop as the entanglement drop, we can equivalently consider the entropy drop of the radiation system due to the small amount of new radiations emitted which are highly entangled with the large system (in this case it is the early radiation) assuming the Haar randomness. Then the Page curve is symmetrical since the von Neumann entropy change of a system with an additional certain amount of thermal radiation is the minus of the system subtracting the thermal radiation. One can argue that the radiation process is irreversible and the time reversal symmetry is not valid. However, the irreversiblity applies to the thermodynamic quantities and has nothing to do with the von Neumann entropy, plus the radiation has been calculated to be thermal in the evaporation. Besides, the von Neumann entropy is always the same for both the radiation and the black hole whether or not they are in equilibrium. Therefore, we have a seemingly contradictory result. Technically, this is due to the ambiguity concerning whether a small amount of radiation just emitted from the black hole can be included in the averaging in Page's theorem, or equivalently, whether the thermal entropy of the small amount of late radiation just emitted can be viewed as entanglement entropy with the large system (early radiation) or it is the black hole area that should be identified as the fine-grained entropy. The purported contradiction of the two arguments is related to the paradox of the radiation density matrix at late times due to our ignorance about the state of the radiation.

\section{Summary and discussion}

Qualitatively, the results can be roughly understood as following: the whole parameter family as different theories can be identified through field redefinitions which leave the island contributions to the entanglement entropy invariant. The CFT part depends on the specific geometries but the late-time and large-distance trend is roughly the same. Therefore, the Page curves are similar for the whole parameter family and the results should be compared with the RST model. The main differences are the following. In our study here, the class of solutions covers black holes of different geometries and topologies which include the RST black hole with singular boundaries and also topologically different models such as the BPP black hole that has an infinite throat approaching the spacetime boundary and is geodesic complete at the end of the evaporation. This extends the results of the Page curve to large range of solutions. Besides, in the late time and large distance limits, the differences in the geometries are washed away and we find a unified answer for the entanglement entropy for the whole family. In addition, a distinct effect shows up only when we smoothly glue the two geometries together and look at its influence on the islands and radiation entropy. The gluing, which is smooth along the gluing surface, causes the island configurations to change abruptly. Consequently, scenarios that demonstrate a second Page transition point show up. Finally, the Page curves are found to be asymmetric in the beginning and end of the evaporation process though the black hole radiates at a constant rate and temperature. This may cause issues related to the purification of early radiation and the state of radiation according to Page’s theorem.

The results for the eternal and evaporating cases are summarized as follows. The boundary of the island in an eternal RST-BPP black hole locates slightly outside the event horizon, which is consistent with many other eternal black hole models, e.g. \cite{Almheiri:2019yqk,Hashimoto:2020cas,Wang:2021woy,Anegawa:2020ezn}. The initial period of the evaporation process sees a linear growth of the entanglement entropy, which suggests that Hawking's calculation gives the correct fine-grained entropy in this period. The entropy is only due to the 2D matter CFT outside the cutoff surface (point). After the Page time, the contribution to the entropy is mainly from the area of the island. The entropy of the black hole is bounded by the Bekenstein-Hawking entropy. For the dynamical black holes formed by collapsing matter shells, the island is inside the event horizon and the Page transition occurs at a third of the evaporation time. The fine-grained entropy of the black hole starts to decay at the half rate it grows when it hits the maximal value of two-thirds the Bekenstein-Hawking entropy $\frac{2}{3}S_{\rm BH}$. Therefore, the Page time occurs at a third of the black hole lifetime. The island hits the singularity slightly before the endpoint of the evaporation, the difference is negligible for massive black holes and does not influence our analysis. 

For the glued black holes geometries which are initially in equilibrium with the incoming radiations emitted from external sources and then start to evaporate, the island may appear in both regions of the spacetime---the equilibrium black hole region and the bath-removed region, or only appear in the bath-removed region depending on the time when the bath is removed. If the entropy of the black hole has already reached the Bekenstein-Hawking entropy before the removal of the bath is sensed by the black hole, the island will jump from outside the horizon to the inside of the event horizon following the null rays of the backreaction of the bath removal. The discontinuity of the island together with the continuous change of entanglement entropy suggests a large amount of entropy near the horizon. The glued geometry admits a ``plateaued" Page curve, which has two different Page transitions. The transition point which shows up in an ordinary evaporating Page curve is separated into two in this model. However, exactly what happens at the two transitions so that the entropy starts to cease increasing or to decrease is unclear microscopically besides using the language of islands.

The drop of the entanglement entropy at late times is slower than the initial increase, as shown in Fig.~\ref{fig:page_gen_removal}. This may not seem to be a problem at first sight but this raises the issue of the purification rates at late times between the Bekenstein-Hawking entropy and the Page's theorem. One resolution can be that the island calculation is only an upper bound of the Page curve and does not represent how the entropy actually behaves. Including more islands is equivalent to summing up more complicated topologies in the corresponding replica manifolds. Though how to count the contributions of all the saddles is not clear from the gravitational Euclidean path integral, but these higher topologies are suppressed by their genus numbers and it is reasonable to believe that their contribution is small. The other resolution is that the quantum informational interpretation is incorrect while the Bekenstein-Hawking entropy is accurate independent of whether the black hole is in equilibrium or not. However, the purity of the final stage of evaporation indicates some mechanisms of purification exist. This contradiction is directly related to the state paradox of the radiation \cite{Akers:2019nfi,Bousso:2020kmy}. The island calculation gives the answer for the entanglement entropy but does not gives details about the purification or how the information in the black hole comes out from the radiation.

In our analysis, we assumed that the black hole is massive and entropy of a single-sided black hole can be approximated as half of that in the double-sided black holes. Our results rely on the large-N limit and the conformal symmetry of the matter fields. The scrambling time can be calculated from the ``dropping" experiment, which counts the time a beam of ingoing light near the horizon takes to reach the entanglement island. This gives $S_{\rm scr}=\frac{\beta}{2\pi}\log \frac{S_{\rm BH}}{N}$ to the leading order, which is approximately the time it takes for the information to come out from an old black hole in terms of radiation \cite{Sekino:2008he,Hayden:2007cs}. Since the boundaries of the islands for massive black holes after Page time show up close to the event horizon, the scrambling time takes the same form for a wide range of the black holes \cite{Almheiri:2019yqk,Hashimoto:2020cas,Wang:2021woy,Hartman:2020swn,Hollowood:2020cou}. However, one should note that the interpretations of the island and of the formula of the quantum extremal surfaces are still far from being unanimous apart from the above-mentioned issue of purification rates. For example, the position of islands are found to be in strange locations in some asymptotically-flat or de Sitter spaces \cite{Li:2021lfo,Sybesma:2020fxg}. Besides, it was argued that the information paradox is resolved at the cost of a new state paradox \cite{Akers:2019nfi,Bousso:2020kmy}. Ideas such as the gravity/ensemble duality or that the incorrect state was used in the computation which are proposed to resolve the state paradox are not yet clearly understood and are not discussed in this study \cite{Bousso:2020kmy,Bousso:2019ykv}. Further studies are needed to deepen our understanding of the interior of the black holes and shed light on our understanding of the entanglement in quantum gravity.

\acknowledgments
X.W wants to thank Kun Zhang for discussions of several related issues on quantum information, Watse Sybesma for the discussions of the interpretation of the Page curve in RST model, and Xinyuan Li for discussions of conformal geometry.

\end{document}